\documentclass[twocolumn,showpacs,preprintnumbers,amsmath,amssymb]{revtex4}
\usepackage{graphicx}
\usepackage{dcolumn}
\usepackage{bm}
\usepackage{graphics}
\usepackage{psfrag}
\usepackage{epsfig}
\usepackage{amssymb}

\begin{document}

\title{Spatial correlation of solar wind fluctuations and solar cycle dependence}

\author{R. T. Wicks}

\author{S. C. Chapman}

\author{R. O. Dendy}

\altaffiliation[Also at]{UKAEA Culham Division, Culham Science Centre, Abingdon, Oxfordshire, OX14 3DB, UK.}
\affiliation{Centre for Fusion, Space and Astrophysics, Physics Department, University of Warwick, Coventry, CV4 7AL, UK.}

\date{\today}

\begin{abstract}
We investigate the spatial correlation properties of the solar wind using simultaneous observations by the ACE and WIND spacecraft. We use mutual information as a nonlinear measure of correlation and compare this to linear correlation. We find that the correlation lengthscales $\lambda$ of fluctuations in density $\rho$ and magnetic field magnitude $|B|$ vary strongly with the solar cycle, whereas correlation lengths of fluctuations in B field components do not. We find $\lambda(|B|) \sim 120 R_E$ and $\lambda(|B|) \sim 270 R_E$ at solar minimum and maximum respectively and $\lambda(\rho) \sim 75 R_E$ and $\lambda(\rho) \sim 170 R_E$ at minimum and maximum. The components of the B field have $\lambda \sim \lambda(|B|)$ at minimum.
\end{abstract}

\maketitle

\section{Introduction}
\textit{In situ} solar wind plasma observations over the last 30 years show that its local properties at 1 a.u. are modulated by the solar cycle \cite{HapgoodSWV} and that this has an impact on the magnetosphere and ionosphere \cite{Johnson}. There is evidence for more mixing of fast and slow solar wind plasma at solar maximum \cite{Bame} and changing magnetic for complexity over the cycle \cite{Kiyani} \cite{HnatGRL}. One method for quantifying the effect of solar activity on the solar wind is to determine the correlation length scale of fluctuations measured \textit{in-situ}. In the past this was done using the Taylor hypothesis \cite{Taylor} and long timeseries from single spacecraft (e.g. Bruno \textit{et al} \cite{Bruno}). More recently, multiple spacecraft studies \cite{Matthaeus} have used linear correlation to calculate a typical correlation length for magnetic field in the solar wind. The ACE and WIND spacecraft give an excellent range of separations and enable long baseline correlations to be calculated for magnetic field vector \textbf{B} and magnitude $|B|$ as well as solar wind bulk velocity and density $\rho$. These observations have been used to show anisotropy in the solar wind \cite{Milano} and measure the Taylor microscale \cite{Matthaeus}. Quantitative knowledge of the spatial correlation of the fluctuations of the flow is relevant to attempts to understand the solar wind in terms of locally evolving turbulence \cite{Goldstein} \cite{Veltri}. It also has implications for cosmic ray propagation  \cite{Zank} \cite{Parhi} which shows solar cycle dependence \cite{Minnie}.
\par
Previous studies used linear correlation measures and did not investigate solar cycle effects on the measurements of correlation length. Here we use a nonlinear measure of correlation, mutual information \cite{Johnson} \cite{Shannon} \cite{TKMarch} \cite{Sello} \cite{Wicks}, alongside linear correlation coefficient. We calculate correlation lengths for the components of $\textbf{B}$, as well as $|B|$ and $\rho$. We also investigate the behaviour of the correlation length of these quantities over the recent solar cycle, using ACE and WIND spacecraft data.

\section{Measuring Correlation}

We use two methods for estimating correlation: correlation coefficient as a linear measure, and normalised mutual information as a nonlinear measure. The linear cross covariance provides a measure of correlation between two signals $A$ and $B$, defined by:

\begin{eqnarray}
C(A,B) = \frac{E[(A - \overline{A})(B - \overline{B})]}{\sqrt{E[(A - \overline{A})^2] E[(B - \overline{B})^2]}} \label{eq:XCor}
\end{eqnarray}
where $E[\ldots]$ denotes the mathematical expectation value and $\overline{A} = E[A]$. Mutual information (MI) quantifies the information content shared by two signals $A$ and $B$. For discrete signals we can write the MI as:

\begin{equation}
I(A,B) = \sum_{i,j}^m P(a_i, b_j) \log_2 \left( \frac{P(a_i, b_j)}{P(a_i)P(b_j)} \right) \label{eq:MI1}
\end{equation}

Here the signals $A$ and $B$ have been partitioned into an alphabet (an exhaustive discrete set which spans the possible values the signal can take) so that $A = \{a_1, \ldots, a_i, \ldots a_m\}$ where $a_1$ and $a_m$ are the extrema of $A$ found in all data considered. The discretised signal takes value $a_i$ with probability $P(a_i)$ and similarly for $b_i$ we have $P(b_i)$, while $P(a_i, b_j)$ is the joint probability of $a_i$ and $b_j$. The chosen base of the logarithm defines the units in which the MI is measured. Normally base 2 is used, so that the MI is measured in bits. If one defines the entropy of a signal as
\begin{eqnarray}
H(A) & = & -\sum_{i}^m P(a_i)\log_2(P(a_i)), \label{eq:MI2}
\end{eqnarray}
then MI can be written as a combination of entropies \cite{Shannon}

\begin{eqnarray}
I(A,B) & = & H(A) + H(B) - H(A,B) \label{eq:MI3}
\end{eqnarray}

The calculation of the entropies needed to form the MI is not trivial, as there is some freedom in the method of discretisation of the signals and in the method used to estimate the probabilities $P(a_i)$, $P(b_j)$ and $P(a_i,b_j)$. There are many different methods currently used, summarised and compared by Cellucci \textit{et al.} \cite{Cellucci} and Kraskov \textit{et al.} \cite{Kraskov}. For spacecraft observations of the solar wind, we use a discretisation based on the standard deviation $\sigma$ of the data. Only data within $5\sigma$ of the mean is considered and bins of $\frac{1}{2}\sigma$ are used, giving $20$ bins in total. The form of MI used here is the normalised mutual information (NMI) \cite{Studholme}, which is $I(A,B)$ normalised by the joint entropy $H(A,B)$ so as to remove the dependence on the entropy of the solar wind at the time of observation:
\begin{eqnarray}
NMI(A,B) & = & \frac{H(A) + H(B)}{H(A,B)} - 1 \label{eq:NMI}
\end{eqnarray}
This gives results in the range $0 \leq NMI \leq 1$ facilitating direct comparison between different periods of data.

\section{Results}

\subsection{The Datasets}

To investigate the effect of the solar cycle on spatial correlation in the solar wind we take data from periods as near to solar activity minimum as possible, that is in 1998, 2005 and 2006, and from solar maximum in 2000. In 1998 WIND was returning towards Earth from the Sun-Earth libration point while ACE orbited the libration point; in 2005 and 2006 both ACE and WIND orbit the libration point. 17 periods of data were chosen with different spacecraft separations, giving a total of 48 days of data. At solar maximum in 2000 WIND is orbiting the Earth and therefore only short periods, when the spacecraft is suitably far from the magnetopause and bow shock, can be used. Four such periods were used giving a total of 15 days of data. Two minute cadence data are used for all variables. We use two different lengths $\tau$ for our measurement window to optimise for the two distinct timescales in the power spectrum. We choose $\tau_L = 960$ minutes to investigate the large scale structures and $\tau_S = 200$ minutes to access the inertial range of turbulence in the solar wind. A shorter window cannot be used as the data in the window then becomes too short for a reliable correlation estimate to be made. NMI (\ref{eq:NMI}) and correlation coefficient (\ref{eq:XCor}) are calculated for windows $\tau_S$ and $\tau_L$ as they are moved along the data. A value for NMI and correlation coefficient is thus obtained every two minutes for each window length, 24 hour averages are calculated and plotted against separation in Figures \ref{fig:Figure1} to \ref{fig:Figure4}.

\subsection{Analysis}

\begin{figure}[ht]
\includegraphics[width=9cm]{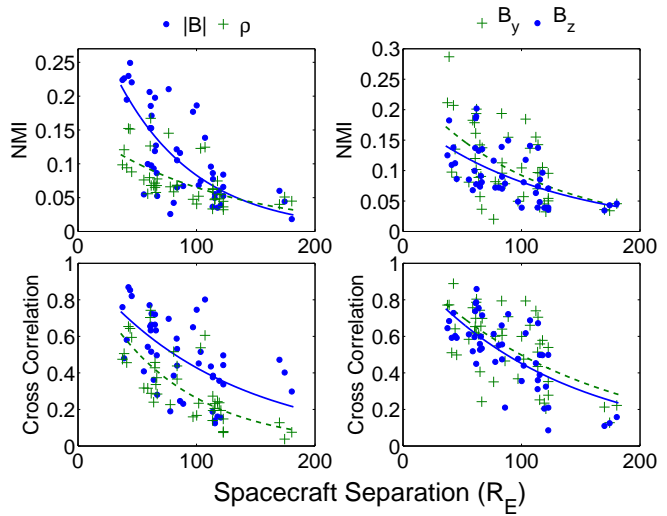}
\caption{NMI and linear correlation for $\tau_L = 960$ minutes, 24 hour averages plotted here versus spacecraft separation. Density $\rho$ ((+) and dashed line) is compared to magnetic field magnitude $|B|$ (($\bullet$) and solid line) in the two left hand panels. In the right hand panels the \textit{y} and the \textit{z} components (GSE) of $\textbf{B}$ are compared. The lines are inverse exponential fits to the data.}
\label{fig:Figure1}
\end{figure}

Figure \ref{fig:Figure1} plots linear correlation and NMI versus spacecraft separation for $\rho$ , $|B|$ and $\textbf{B}$ components at solar minimum using $\tau_L$. Following \cite{Matthaeus} we fit an exponential function $y = a \exp (x/\lambda)$ to the data using nonlinear least squares, with $a = 1$ for correlation coefficient. The values of $\lambda$ calculated are shown in Table \ref{table:Table1}. The quality of the fit is indicated by the error on the correlation length, calculated as the $95\%$ confidence bound of the nonlinear least squares fit. The errors on these values arise primarily from the scatter of the measurements due to the highly variable nature of the solar wind.
\par

\begin{table}

\begin{tabular}{ccccc}
\hline
\multicolumn{1}{c}{Variable} & \multicolumn{2}{c}{Normalised MI} & \multicolumn{2}{c}{Correlation Coeff.} \\ 
& $\lambda_{min}$ & $\lambda_{max}$ & $\lambda_{min}$ & $\lambda_{max}$ \\ 
\hline
$|B|$ & $67 \stackrel{+31}{_{-16}}$ & $219 \stackrel{+126}{_{-106}}$ & $118 \stackrel{+21}{_{-15}}$ & $274 \stackrel{+104}{_{-59}}$ \\
$\rho$ & $115 \stackrel{+99}{_{-37}}$ & $^*$ & $75 \stackrel{+11}{_{-9}}$ & $167 \stackrel{+78}{_{-40}}$ \\
$B_y$ & $102 \stackrel{+102}{_{-34}}$ & $^*$ & $144 \stackrel{+16}{_{-18}}$ & $169 \stackrel{+40}{_{-27}}$ \\
$B_z$ & $117 \stackrel{+103}{_{-37}}$ & $^*$ & $125 \stackrel{+66}{_{-31}}$ & $169 \stackrel{+34}{_{-24}}$ \\
\hline
\end{tabular}

\caption{Values of correlation length $\lambda$ using window $\tau_L$, with $95\%$ confidence bounds, in Earth radii, calculated by least squares fitting of exponentials to the 24 hour mean NMI and linear correlation coefficient during solar minimum (1998, 2005, 2006) and solar maximum (2000). Those values marked with $^*$ are fits that resulted in an $R^2$ value of less than $0.1$.} 
\label{table:Table1}
\end{table}

\par
We anticipate the spatial variation of correlation coefficient to be of exponential form. However NMI, which is a nonlinear measure, may have more complex behaviour; an exponential function provides a simple parameterisation of the data. The NMI gives $\rho$ a longer correlation length than $|B|$, whereas the correlation coefficient gives the reverse. Both give a lower correlation for $\rho$ than $|B|$. If both $\rho$ and $|B|$ NMI fits are set to have the same intercept then the NMI gives similar results to those found with the cross corelation, however as exact form of the NMI dependence on separation is unknown these correlation lengths remain a local estimate.
\par
Table \ref{table:Table1} also shows that at minimum the \textit{y} and \textit{z} components of $\textbf{B}$ have similar correlation lengths to each other and $|B|$. The scatter on the correlation calculated for the components of $\textbf{B}$ and $|B|$ is larger than for $\rho$. The correlation coefficient gives a clearer decay and the estimates of correlation length are correspondingly more accurate.
\par

\begin{figure}
\includegraphics[width=9cm]{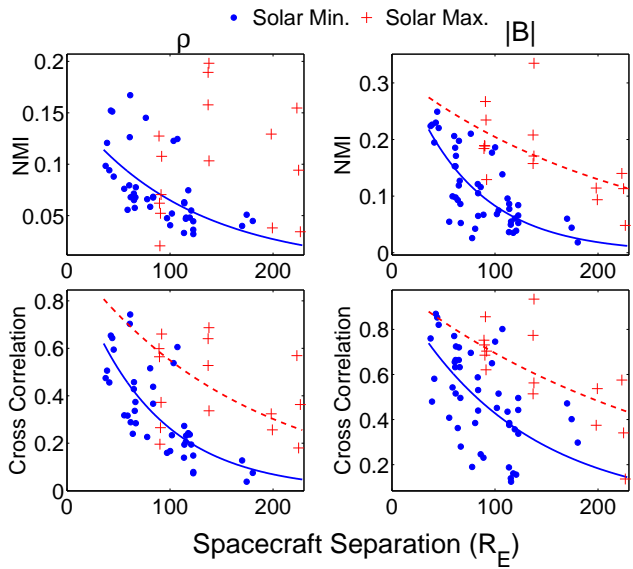}
\caption{24 hour mean NMI and correlation coefficient coefficients, between ACE and WIND $|B|$ and $\rho$ for window length $\tau_L = 960$ minutes, for selected periods in 1998, 2005 and 2006 (solar minimum: ($\bullet$) and solid line) and 2000 (solar maximum: (+) and dashed line). The lines plotted are inverse exponential fits to the data.}
\label{fig:Figure2}
\end{figure}

\begin{figure}
\includegraphics[width=9cm]{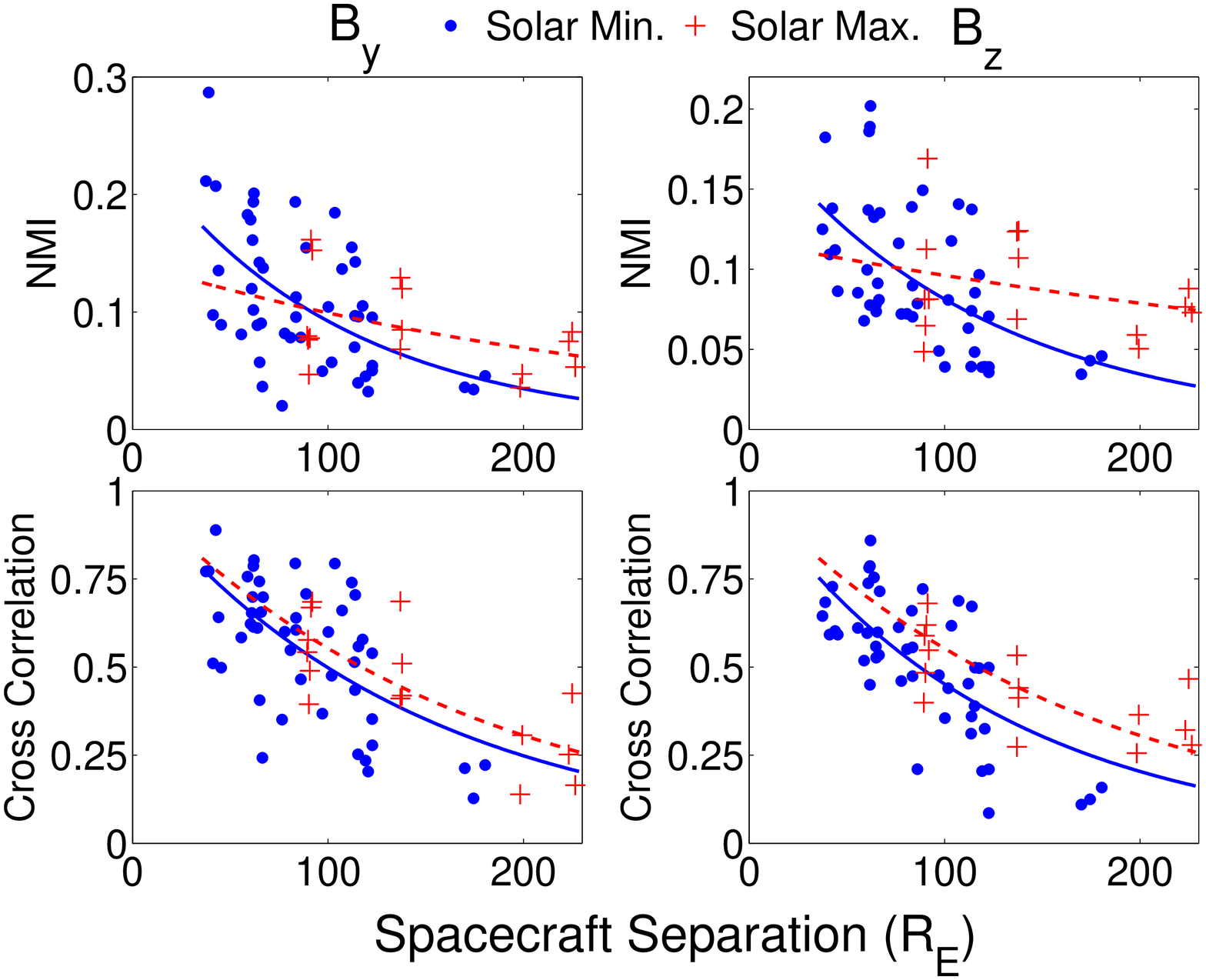}
\caption{24 hour mean NMI and correlation coefficient coefficients, between ACE and WIND $B_y$ and $B_z$ using window length $\tau_L = 960$ minutes, for selected periods in 1998, 2005 and 2006 (solar minimum: ($\bullet$) and solid line) and 2000 (solar maximum: (+) and dashed line). The lines plotted are inverse exponential fits to the data}
\label{fig:Figure3}
\end{figure}

We now consider variation with solar cycle. Figure \ref{fig:Figure2} plots linear correlation coefficient and NMI against spacecraft separation for $|B|$ and $\rho$. It shows that both the linear and the nonlinear measure of correlation yield a higher spatial correlation in both $\rho$ and $|B|$ at solar maximum. To quantify this we again fit an exponential of the form $y = a \exp (x/\lambda)$, with $a = 1$. This provides a reasonable fit in all cases except for the NMI at solar maximum for $\rho$, which also shows the largest scatter. The calculated values of correlation length $\lambda$ are shown in Table \ref{table:Table1}. We see that the correlation length measured at solar minimum is systematically smaller than that at maximum. The value at maximum for $|B|$ is within error of previous estimates for the same interval \cite{Matthaeus}.
\par
Figure \ref{fig:Figure3} shows the spatial correlation of components of $\textbf{B}$. The points from solar maximum and minimum do not show distinct behaviour, unlike $\rho$ and $|B|$ in Figure \ref{fig:Figure2}. Fitting an exponential function to the data, as before, yields the values of the correlation lengths for the components of $\textbf{B}$ shown in Table \ref{table:Table1}. The $B_y$ and $B_z$ values of $\lambda_{max}$ and $\lambda_{min}$ are within error of each other, and are distinct from the values for $|B|$ and $\rho$.

\begin{figure}
\includegraphics[width=9cm]{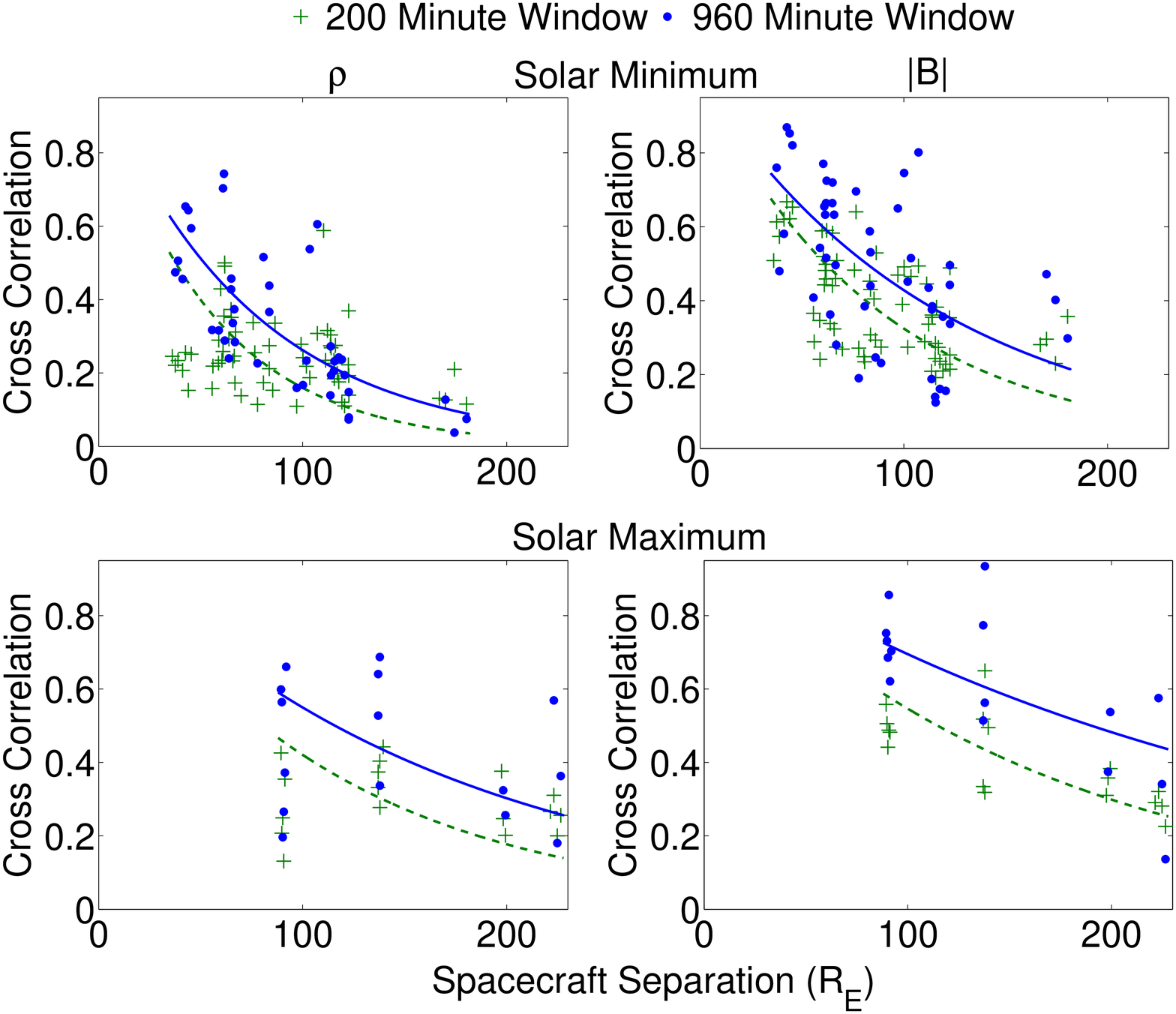}
\caption{24 hour mean correlation coefficient coefficients, calculated between ACE and WIND $|B|$ and $\rho$ measurements at solar minimum in 1998, 2005 and 2006 (top panels)  and solar maximum in 2000 (bottom panels). The ($\bullet$) and solid line fits correspond to $\tau_L = 960$ min and the (+) and dashed line fits correspond to $\tau_S = 200$ min. The lines are inverse exponential fits to the data.}
\label{fig:Figure4}
\end{figure}

Finally, Figure \ref{fig:Figure4} gives a comparison between the measured correlation coefficient from the two different time windows, $\tau_L = 960$ minutes and $\tau_S = 200$ minutes. The shorter window is within the inertial range of solar wind turbulence; it gives slightly smaller values for the correlation length ($\lambda_{min}(|B|) =  89 \stackrel{+8}{_{-7}} R_E$, $\lambda_{min}(\rho) =  55 \stackrel{+7}{_{-6}} R_E$) than the longer time window ($\lambda_{min}(|B|) = 118 \stackrel{+21}{_{-15}}$, $\lambda_{min}(\rho) = 75 \stackrel{+11}{_{-9}}$). This difference is more obvious at solar maximum where the correlation lengths measured by the smaller window are $\lambda_{max}(|B|) = 166 \stackrel{+26}{_{-21}} R_E$ for $|B|$ and $\lambda_{max}(\rho) = 116 \stackrel{+35}{_{-22}} R_E$ for $\rho$; compared to $\lambda_{max}(|B|) = 274 \stackrel{+104}{_{-59}} R_E$ and $\lambda_{max}(\rho) = 167 \stackrel{+78}{_{-40}} R_E$ using the longer window $\tau_L$. Large coherent structures within the solar wind may contribute to the change in correlation length seen on timescales $\tau_L$ above the inertial range value $\tau_S$; however the difference is not large.

\section{Conclusions}
We have used simultaneous data from the ACE and WIND spacecraft separated by between $30$ and $220 R_E$, to calculate the spatial correlation of solar wind $\rho$, $|B|$ and \textbf{B} components. Cross correlation and normalised mutual information were used as alternative linear and nonlinear measures of correlation. At solar minimum we have $48$ days of contemporaneous WIND and ACE observations. We determine the correlation lengthscale $\lambda(\rho)$ for $\rho$ to be $\lambda(\rho) = 75 R_E$. This is smaller than $\lambda(|B|)$ by $\sim$ 1.6. For the components of \textbf{B}, we find $\lambda(B_z)$ is within error of $\lambda(|B|)$ and $\lambda(B_y)$ is slightly above the range of errors.
\par
At solar maximum we have $15$ days of contemporaneous WIND and ACE data. This allows us to investigate the effect of the solar cycle. We find that $\lambda(|B|)$ and $\lambda(\rho)$ at solar maximum are larger than at minimum by $\sim$ a factor of $2$. In contrast the components of the magnetic field show weak variation with solar cycle.
\par
We have used two window lengths; one within the inertial range ($200$ min) and the other on longer timescales ($960$ min).The window on inertial range timescales gives values of $\lambda$, that are systematically shorter, but within the errors, of the values from the larger window. The solar cycle dependence of the correlation length is in all cases independent of our chosen window sizes.
\par
Our result that $\rho$ and $|B|$ show variation with solar cycle, whereas the components of \textbf{B} do not, is consistent with the idea that the correlation in $|B|$ and $\rho$ is, at least in part, of solar origin. Intriguingly this behaviour persists when we restrict our analysis to timescales within the inertial range. This is consistent with recent single-spacecraft results \cite{Kiyani} \cite{HnatGRL} which show solar cycle dependence within the inertial range of magnetic energy density fluctuations.

\section{Acknowledgements}
The authors acknowledge the STFC, EPSRC and UKAEA for support, the CSC (Warwick) for computing facilities, the WIND MFI and SWE teams and the ACE MAG and SWEPAM teams for data provision.

 \end{document}